\documentclass[aps,pra,superscriptaddress,amsmath,amssymb,preprintnumbers,twocolumn,showpacs]{revtex4}
\usepackage{amssymb}
\usepackage{color}
\usepackage{graphicx}
\usepackage{subfigure}

\definecolor{dgreen}{rgb}{0,0.5,0}

\begin{document}
\def\bbm[#1]{\mbox{\boldmath$#1$}}

\title{Reconstruction of time-dependent coefficients: a check of approximation schemes for non-Markovian convolutionless dissipative generators}

\author{Bruno Bellomo}\affiliation{MECENAS, Universit\`{a} Federico II di Napoli \& Universit\`{a} di Bari, Italy}\affiliation{CNISM and
Dipartimento di Scienze Fisiche ed Astronomiche, Universit\`a di
Palermo, via Archirafi 36, 90123 Palermo, Italy}

\author{Antonella De Pasquale}\affiliation{MECENAS, Universit\`{a} Federico II di Napoli \& Universit\`{a} di Bari, Italy}\affiliation{Dipartimento di Fisica, Universit\`{a} di Bari, via Amendola 173, I-70126 Bari, Italy; \\ INFN, Sezione di Bari, I-70126 Bari, Italy}

\author{Giulia Gualdi}\affiliation{MECENAS, Universit\`{a} Federico II di Napoli \& Universit\`{a} di Bari, Italy}\affiliation{Dipartimento di Matematica e Informatica, Universit\`a degli Studi di Salerno,
Via Ponte don Melillo, I-84084 Fisciano (SA), Italy; CNR-INFM
Coherentia, Napoli, Italy; CNISM Unit\'a di Salerno; INFN Sezione di
Napoli gruppo collegato di Salerno, Italy}

\author{Ugo Marzolino}\affiliation{MECENAS, Universit\`{a} Federico II di Napoli \& Universit\`{a} di Bari, Italy}
\affiliation{Dipartimento di Fisica, Universit\`{a} di Trieste,
Strada Costiera 11, 34151, Trieste, Italy;\\ INFN, Sezione di
Trieste, 34151, Trieste, Italy}

\begin{abstract}

We propose a procedure to fully reconstruct the time-dependent coefficients of  convolutionless non-Markovian dissipative generators via a finite number of experimental measurements. By  combining a tomography based approach  with a proper data sampling, our proposal allows to relate the time-dependent coefficients governing the dissipative evolution of a quantum system to experimentally accessible quantities.  The proposed scheme not only  provides a way to retrieve full information about potentially unknown dissipative coefficients but also, most valuably, can be employed as  a reliable consistency test for the approximations involved in the theoretical derivation of a given non-Markovian convolutionless master equation.
\end{abstract}


\maketitle

\section{Introduction}

 The dissipative evolution of a quantum system interacting with an environment represents a phenomenon of paramount importance in quantum information science and beyond, as it addresses a fundamental issue in quantum theory.
In general a complete microscopic description of the dynamical evolution of a system  coupled to the environment (or bath) is a complex many-body problem which requires the solution of a potentially infinite number of coupled dynamical equations. According to an open system approach, this issue is  tackled by retaining only basic information about the environment and describing  the system dynamics in terms of a master equation \cite{Petruccione-Breuerlibro2002, Gardiner-Zoller}. The lack of a complete knowledge about the bath leads to master equations coefficients (MECs) which may be either unknown, or  obtained from a microscopical derivation carried out within some approximation scheme. As a matter of fact it would then be highly appealing to devise a procedure allowing to retrieve the MECs starting from experimentally accessible quantities.  This would in fact  both give access to otherwise unknown quantities and provide a strong indication about the validity of the adopted theoretical framework.

So far two main dynamical regimes, Markovian and non-Markovian, can usually be distinguished according to the timescale of environment dynamics (respectively shorter or longer than that of the system).  In \cite{Bellomo2009} it has been shown that, in case of Markovian Gaussian noise,  the  MECs can be retrieved by means of a finite number of tomographic measurements by using Gaussian states as a probe.
Here we  want to address the more involved non-Markovian case. In facts, even though Markovian evolutions  have been extensively investigated (see e.g. \cite{Petruccione-Breuerlibro2002, Gardiner-Zoller, BenattiFloreanini2005, Spohn1980}),  in general real noisy dynamics are far from being Markovian. Despite a growing interest in both theory and experiment  \cite{Hope2000,Pomyalov2005} a comprehensive theory of non-Markovian dynamics is yet to come.
Exact non-Markovian master equations have been derived for a Brownian particle linearly coupled to a harmonic oscillator bath via e.g. path integral methods \cite{HPZ1992,HM1994} or phase-space and Wigner function computations \cite{HR1985,HY1996}.  Analogous results have been obtained employing quantum trajectories, either exactly or in  weak coupling approximation \cite{Diosi1997,Strunz2004,Yu2004,Bassi2009}. In the framework of path integral methods, master equations have been derived both for initially correlated states \cite{KG1997,RP1997}, and for factorized initial states in the case of weak non linear interactions \cite{HPZ1993}. Nevertheless,
all these master equations cover only few cases and are not simple to solve. Indeed, it would be highly desirable to find an approximation scheme fully capturing non-Markovian features, as
in general different approximations may  lead to irreconcilable dynamics \cite{Petruccione-Breuerlibro2002}.

 When dealing with a time-dependent generator one can face two, in principle distinct, kinds of problems. On one hand, taking for granted the functional form of the dissipative generator, one might be interested in retrieving the time-independent parameters which characterize it. This issue has been tackled in  \cite{Bellomo2010}, were it has been presented a  first generalization of the tomography-based approach proposed in \cite{Bellomo2009} to time-dependent dissipative generators.  There it has been shown that, under the assumption of Gaussian noise, the time-independent parameters  (TIPs) characterizing the time-dependent MECs (whose functional form is previously known) as for example the system-bath coupling, temperature and bath frequency cut-off, can be obtained with a finite set of measurements. On the other hand, one might be interested in the more general problem of  reconstructing the functional form of the dissipative generator itself. This might be the case  either because the time-dependence of the dissipative generator  is completely unknown or because one wants to test the validity of the theoretical assumptions at the basis of the dissipative model.

In this paper we propose an experimentally feasible procedure which allows
the full reconstruction of the  time-dependence of convolutionless non-Markovian generators. Indeed, whenever the assumption of Gaussian noise is satisfied, by combining a tomography based approach and a proper choice of data sampling, starting from a finite and discrete set of measurements it is possible to obtain global information about the time-dependent MECs. We envisage two main applications of the proposed  scheme: both a full reconstruction starting from previously unknown MECs, and a sound, reliable and complete consistency test of the theoretical assumptions made in deriving the dissipative generator.

The paper is organized as follows. In section \ref{TC} we review the  tomography based approach introduced in \cite{Bellomo2009}  allowing to reconstruct the first two cumulants of a Gaussian state at any fixed time with a finite amount of measurements. In section \ref{sec:MEPSreconstruction} we introduce the main lines of our reconstruction  scheme for two in principle distinct cases based on the purpose of the reconstruction and on the available prior knowledge of the MECs. In section \ref{sec:approachs} two different implementations, integral and differential, of the reconstruction procedure are presented and their application is discussed. To provide an explicit example of both approaches, in section \ref{par:specific case} they are applied to the specific model of a brownian particle interacting with an Ohmic bath of harmonic oscillators. In section \ref{par:Summary and Conclusions} we summarize and discuss our results, and finally in Appendix \ref{sampling} we provide some more details about the sampling theorems involved in our reconstruction scheme.

\section{The T-C procedure } \label{TC}

In this section we briefly review the main steps of the procedure introduced in \cite{Bellomo2009} which allows to relate the evolved first and second order cumulants of a Gaussian state to tomographic measurements.
We consider a master equation with unknown coefficients which generates a Gaussian Shape Preserving (hereafter GSP) dissipative evolution or, in other words, a Gaussian map. We then focus on the time evolution of the first and second cumulants of a Gaussian state, which completely determine its dynamics. Given a master equation governing the evolution of the density matrix of the system $\hat\rho(t)$,  the dynamical equations for the  first two cumulants formally read
\begin{eqnarray}\label{cumulants}
\langle \hat{q}\rangle_t &=& {\rm Tr}(\hat{\rho}(t)\hat q), \nonumber \\
\langle \hat{p}\rangle_t &=& {\rm Tr}(\hat{\rho}(t)\hat p),\nonumber \\
\Delta q_t^2 &=& {\rm Tr}(\hat{\rho}(t)\hat q^2)-\langle \hat{q}\rangle^2_t, \nonumber \\
\Delta p_t^2 &=&  {\rm Tr}(\hat{\rho}(t)\hat p^2)-\langle \hat{p}\rangle^2_t, \nonumber \\
\sigma(q,p)_t & = & {\rm Tr}\left(\hat{\rho}(t)\frac{\hat{q}\hat{p}+\hat{p}\hat{q}}{2}\right)-\langle \hat{q}\rangle_t\langle \hat{p}\rangle_t.
\end{eqnarray}

The explicit form of these equations depends on the adopted master equation whose coefficients, in general, are either unknown or derived by means of phenomenological assumptions.
 Inserting  into Eqs.~(1) the explicit expression of $\hat{\rho}(t)$ and then solving for the MECs,  allows to write these coefficients (or the differential equations they satisfy) as a function of  the first two cumulants of a Gaussian probe. Hence, by measuring these two quantities, one can gain indirect experimental information about the MECs.
To this aim, symplectic quantum tomography arises naturally as a key tool to perform the required measurements. In particular we will use the symplectic transform, or $M^2$-transform \cite{manko2000}, that finds its natural implementation in experiments with massive particles, as for example those to detect the longitudinal motion of neutrons \cite{GBadurek2006}
and to reconstruct the transverse motional states of helium atoms \cite{Kurtsiefe1997}.
We also note that the symplectic transform is equivalent to the Radon transform \cite{Facchi2010} which is experimentally implemented by homodyne detection in the context of quantum optics \cite{Auria2009}. We now recall a procedure introduced in \cite{Bellomo2009} allowing to connect, at any fixed time $t$, tomographic measurements and the first two cumulants of a Gaussian probe by means of a small number of detections.

We begin by switching from the space of quantum states, i.e. the Hilbert space, to phase-space, which can be done by means of the Wigner map. In facts, a quantum state $\hat{\rho}(t)$  can be represented on phase-space in terms of its Wigner function $W$ defined as
{\begin{equation}\label{wigner function definition}
    W(q,p,t)=\frac{1}{\pi \hslash} \int_{-\infty}^{+\infty}\mathrm{d}y \exp \left(
    \frac{i 2 p y}{\hslash}\right)\hat{\rho} (q-y,q+y,t). \\
\end{equation}
The Wigner function of a Gaussian state is itself a Gaussian function on phase space. Hence, if the dissipative evolution is GSP, at any time the Wigner function of the evolved Gaussian state can be written as
\begin{eqnarray}\label{wigner function gaussian}
W(q,p,t)  &=&  \frac{1}{2 \pi \sqrt{\Delta q_t^2\Delta p_t^2-\sigma(q,p)_t^2}}
  \nonumber\\  &\times & \exp \Bigg[-  \frac{\Delta q_t^2
  (p-\langle \hat{p}\rangle_t)^2+\Delta p_t^2(q-\langle \hat{q}\rangle_t)^2}{2[\Delta q_t^2\Delta p_t^2-\sigma(q,p)_t^2]}\nonumber\\&&\qquad\quad-\frac{2\sigma(q,p)_t (q-\langle \hat{q}\rangle_t)(p-\langle \hat{p}\rangle_t)}{2[\Delta q_t^2\Delta p_t^2-\sigma(q,p)_t^2]} \Bigg]
  \,.
\end{eqnarray}
 In general performing a tomographic map then amounts to project the above Wigner function along a line in phase-space
\begin{equation}\label{line in q p plane}
    X-\mu q - \nu p = 0,
\end{equation}
or, in other words, to compute its symplectic transform

\begin{eqnarray}\label{radon transform}
    \varpi (X,\mu,\nu)&=&\langle \delta\left(X-\mu q - \nu p\right)
    \rangle\nonumber\\ &=&
    \int_{\mathbb{R}^2} W(q,p,t)  \delta\left(X-\mu q - \nu p\right) \mathrm{d}q
    \mathrm{d}p.\nonumber\\
\end{eqnarray}
For the case of Gaussian states, the symplectic transform is again a Gaussian function:
\begin{eqnarray}\label{radon transform gaussian}
& \varpi & (X,\mu,\nu)=\frac{1 }{\sqrt{2\pi}\sqrt{\Delta q_t^2\mu^2+\Delta p_t^2 \nu^2 + 2\sigma(q,p)_t \mu \nu}}\nonumber \\ & \times &  \exp \left[-\frac{\left(X-\mu \langle \hat{q}\rangle_t - \nu \langle \hat{p}\rangle_t \right)^2}{2[\Delta q_t^2\mu^2+\Delta p_t^2 \nu^2 + 2\sigma(q,p)_t \mu \nu]}
    \right],
\end{eqnarray}}
where the second cumulants always obey the constrain $
\Delta q_t^2\mu^2+\Delta p_t^2 \nu^2 + 2\sigma(q,p)_t \mu \nu
>0$ as a consequence of the Schr\"odinger-Robertson relation \cite{Robertson1934},  which represents
a generalization of the Heisenberg principle.
 At this point one could wonder whether, due to experimental errors,  a violation of the
uncertainty principle might be observed. This may
happen if measurements are performed on states almost saturating the
 inequality, i.e. on pure states.
In our case measurements are performed
on states undergoing a dissipative evolution which typically are far from being
pure hence from saturating the uncertainty relation. Furthermore
 any additional noise of statistical origin will
have the effect of moving the reconstructed state further away from the boundary,
as noted in \cite{Rehacek2009}.

The symplectic transform (\ref{radon transform gaussian}) provides a relation between the first and second order cumulants and
the tomogram values along each line.  In facts, performing a tomographic measurement  consists in choosing a pair $(\bar\mu,\bar\nu)$, i.e. a line in phase space, and in computing $\varpi (X,\mu,\nu)$ for a given value of $X$.
In \cite{Bellomo2009} it has been shown that by choosing
the lines  corresponding to
position and momentum probability distributions (i.e. $(\bar\mu,\bar\nu)=(1,0)$ and $(0,1)$) then, for any  fixed $t$,
 at most four points along each line (i.e. tomogram) are needed
to retrieve the first and second cumulant of the associated
variable. Analogously, the covariance of the two variables is obtained by measuring
at most two points along the line $(\bar\mu,\bar\nu)=(1/\sqrt{2},1/\sqrt{2})$. In overall,
starting from a GSP master equation, the first and second
order evolved momenta at time $t$ can be obtained via a total amount of eight or at most ten
points measured in three different directions. In the following, we
will refer to this procedure as the tomograms-cumulants
(T-C) procedure.

  To summarize, at each time $t$, the T-C procedure combined with Eqs.~(\ref{cumulants}) provides
the sought-for bridge between dynamical parameters and measurable quantities. We note that, in the Markovian case, this combination is straightforwardly performed by inverting Eqs.~(\ref{cumulants}) \cite{Bellomo2009}. Finally, we emphasize that  even though the assumption of Gaussian noise might be seen as an idealization, it is actually well fitted for a significant number of models \cite{Petruccione-Breuerlibro2002,Gardiner-Zoller} and small deviations from Gaussianity would only introduce small and controllable errors. Also, Gaussian probes are quite straightforward to produce either with a laser  or  an ordinary light source (obtaining, respectively, a poissonian or a thermal distribution) \cite{Mandel1995}.

\section{Sketch of the reconstruction}\label{sec:MEPSreconstruction}
The T-C procedure allows to gain information about the unknown MECs -via experimentally accessible quantities-  {\it locally} in time. This is enough
 if the  GSP dissipative generator is time-independent,  as the information needed to fully reconstruct  the unknown MECs is obtainable by performing tomographic measurements at one arbitrary instant of time, as shown in \cite{Bellomo2009}.  This is no longer true when facing a time-dependent generator. In this case to fully reconstruct  the unknown MECs we in principle need to gather information {\it globally} in time. Now, taking the T-C procedure as a starting point, the problem we tackle is how to retrieve the full information  we need by combining measurements performed on a finite and discrete set of times, i.e. starting from partial information.
 \subsection{the dissipative generator}
We focus on the reconstruction of  the time-dependent MECs of the following class of master equations

\begin{eqnarray}\label{master equation in q e p}
\frac{\mathrm{d} \hat{\rho}(t)}{\mathrm{d}
t}&=&\!-\frac{i}{\hslash}\left[\hat{H}_0,\hat{\rho}(t)
\right]-\frac{i (\lambda(t) +\delta)}{2
\hslash}\left[\hat{q},\hat{\rho} (t) \hat{p}+\hat{p}\hat{\rho}(t) \right]
\nonumber\\&& +\frac{i (\lambda(t) -\delta)}{2 \hslash}\left[\hat{p},\hat{\rho}(t)
\hat{q}+\hat{q}\hat{\rho}(t) \right]\nonumber\\&&-\frac{D_{pp}(t)}{
\hslash^{2}}\left[\hat{q},[\hat{q},\hat{\rho}] \right]
-\frac{D_{qq}(t)}{ \hslash^{2}}\left[\hat{p},[\hat{p},\hat{\rho}(t)]
\right]\nonumber\\&&+\frac{D_{qp}(t)}{ \hslash^{2}}
\left(\left[\hat{q},[\hat{p},\hat{\rho}(t)]\right]+\left[\hat{p},[\hat{q},\hat{\rho}(t)]
\right]\right),
\end{eqnarray}
where the unknown MECs are $\lambda(t)$, $D_{qq}(t)$, $D_{pp}(t)$, $D_{qp}(t)$.
The system Hamiltonian $H$ is chosen to be at most a second-order polynomial in the position and momentum operators $\hat{p},\hat{q}$ \cite{Sandulescu1987}
\begin{equation}\label{Hamiltonian}
    \hat{H}=\hat{H}_0+\frac{\delta}{2}\left(\hat{q}\hat{p}+\hat{p}\hat{q}\right), \qquad \hat{H}_0= \frac{1}{2m}\hat{p}^2+\frac{m\omega^2}{2}\hat{q}^2,
\end{equation} where the time dependence which may be introduced by the Lamb shift term has been neglected. This is  typically  justified as most of the times either the Lamb shift is negligible or  the Hamiltonian part of the dissipative generator  reaches its asymptotic value on  a much shorter timescale compared to the non unitary part  \cite{Petruccione-Breuerlibro2002}. In overall, the choice of operators in Eq.~(\ref{master equation in q e p}) represents a natural generalization of the GSP time-independent master equation introduced in \cite{Sandulescu1987}. The investigation of this time-dependent class of master equations is further motivated by the existence of a wide range of models obeying a GSP dissipative dynamics of this form
\cite{Strunz2004,Yu2004,Bassi2009,HM1994,HR1985,HY1996,KG1997,RP1997}.

As a side remark we note that there may be some ambiguity in  literature about the relation between time-dependent dissipative generators and Markovian/non-Markovian dynamics. In facts, some authors classify as non-Markovian only  generators containing a convolution integral.  It has recently been proved in \cite{Kossakowski2010} that these generators can be mapped into convolutionless ones. The distinctive feature of non-Markovianity becomes then the dependence of the convolutionless generator on $t-t_0$ where $t_0$ is the initial time. According to this approach a time-dependent generator of the kind (\ref{master equation in q e p}) could be considered as Markovian. However, following a consistent part of literature, e.g. \cite{HPZ1992, HR1985, HY1996, Strunz2004,Yu2004,HPZ1993,Maniscalco2004}, we  will term non-Markovian also convolutionless time-dependent generators as the one in Eq.~(\ref{master equation in q e p}).

\subsection{the reconstruction scheme}

Our strategy towards the reconstruction of the MECs in Eq.~(\ref{master equation in q e p}) is made up of three main steps:
\begin{enumerate}
  \item use the T-C procedure to get indirect measurements of the evolved cumulants at different times;
  \item use these measurements to retrieve the values of the MECs (or functions of them) at those times exploiting the connection between MECs and first cumulants of a Gaussian probe, obtained by employing the dynamical equations (\ref{cumulants});
  \item starting from the obtained discrete and finite set of values, reconstruct the full expression for the MECs by applying proper sampling theorems.
\end{enumerate}
\noindent
In particular, useful for our purposes will be the  Nyquist-Shannon theorem \cite{Jerry1977,Unser2000} and one of its more sophisticated generalizations involving an additive random sampling \cite{Shapiro1960,Beutler1970,Jerry1977}. In principle we can distinguish two different applications of the reconstruction procedure:  check of the {\it a priori} assumed time-dependence of MECs (Case I) or  complete reconstruction  of MECs with no {\it a priori} assumptions (Case II).

{\it Case I:}  we assume {\it a priori} a certain time dependence of the MECs as a consequence, for example, of a microscopical derivation of the master equation. In this perspective, we are interested in experimentally reconstructing the MECs to check the validity of the approximations made. A mismatch between the assumed and the measured MECs would in fact provide a strong evidence of the breakdown of the adopted approximation scheme. In this case a full knowledge of the MECs is assumed, including that of the TIPs involved, such as the bath frequency cut-off, system-bath coupling, etc. The TIPs can be either assumed or previously reconstructed \cite{Bellomo2010}. In this case, given the prior knowledge of the bandwidth associated to the function to be reconstructed  (i.e. the width of its Fourier transform), the suitable sampling theorem can be chosen accordingly.  If  the function is band-limited then  to obtain an exact reconstruction it is enough to apply the simplest sampling theorem, i.e. the Nyquist-Shannon theorem (see Appendix \ref{sampling}). The function can hence be reconstructed starting from a  discrete set of values spaced according to the width of its Fourier transform.  If the bandwidth is not limited one could  truncate it and still apply the same procedure, which would then be affected by the so-called aliasing error. To minimize it one can  perform a proper truncation. Alternatively  a more general additive random sampling, that avoids the aliasing error (see Appendix \ref{sampling}), can be employed.

{\it Case II:} here we want to fully reconstruct the MECs, or derived functions,  with no previous assumption on the dynamics, i.e. the MECs are fully unknown. In general, this implies no prior knowledge of the bandwidth associated to  the function to be reconstructed. In this case we must resort to an additive random sampling (see Appendix \ref{sampling}). If on one hand this procedure involves  function averages with respect to the probability of drawing $n$ sampling times (i.e. more involved measurements), on the other it does not require prior  knowledge of the bandwidth and is an alias-free sampling.  To obtain the averages of the function we should in principle perform measurements over a continuous interval of time, as the reconstruction is proposed with continuous random processes.  In practice, every experimental apparatus employed to record and process the data has a dead working time, such that the random process will be discrete in time, no matter how dense, thus introducing an intrinsic source of error in the procedure.

As a conclusive remark,  we note that in both cases  the set of measurements required turns out to be discrete but in principle infinite, as the reconstruction should be performed over the whole real axis. This number can be made finite by invoking the largely reasonable physical condition of a finite observation time.
\section{Different procedures}\label{sec:approachs}

In this section we illustrate in more detail how to implement the above sketched reconstruction scheme. The starting point are the general equations  (\ref{cumulants}) and (\ref{master equation in q e p}) which  can be expressed in compact matrix form as
\begin{eqnarray}
& & \frac{d}{dt}S(t)=\left(M-\lambda(t) I_2\right)S(t), \label{eq1} \\
& & \frac{d}{dt}X(t)=\left(R-2\lambda(t) I_3\right)X(t)+D(t), \label{eq2}
\end{eqnarray}
where $I_{2(3)}$ is  the 2(3)-dimensional identity matrix. The vectors  $S(t)$ and $X(t)$ correspond, respectively, to the first and second order cumulants
 \begin{eqnarray}
 S(t)=\frac{1}{\sqrt{\hslash}}
\begin{pmatrix}
\displaystyle \sqrt{m\omega}\langle\hat q\rangle_t \\
\displaystyle \frac{\langle\hat p\rangle_t}{\sqrt{m\omega}}
\end{pmatrix}
, \quad X(t)= \frac{1}{\hslash}
\begin{pmatrix}
\displaystyle m\omega(\Delta q)_t^2 \\
\displaystyle \frac{(\Delta p)_t^2}{m\omega} \\
\displaystyle (\sigma_{q,p})_t
\end{pmatrix}
, \;\;\label{momenta}\end{eqnarray}
and the matrices $M$ and $R$ contain the Hamiltonian parameters
\begin{equation} M=
\begin{pmatrix}
\delta & \omega \\
-\omega & -\delta
\end{pmatrix}
, \qquad R=
\begin{pmatrix}
2\delta & 0 & 2\omega \\
0 & -2\delta & -2\omega \\
-\omega & \omega & 0
\end{pmatrix}.\label{MR}\end{equation}
Finally,  $D(t)$ is the diffusion vector
\begin{equation}
D(t)= \frac{2}{\hslash}
\begin{pmatrix}
\displaystyle m\omega D_{qq} (t)\\
\displaystyle \frac{D_{pp}(t)}{m\omega} \\
\displaystyle D_{qp}(t) \\
\end{pmatrix}, \label{D}\end{equation}
such that the MECs to be reconstructed are $\lambda(t)$ and $D(t)$. The dynamical evolution of the first cumulants, Eq.~(\ref{eq1}), only  depends on the friction coefficient $\lambda(t)$ whereas that of  the second cumulants, Eq.~(\ref{eq2}), depends on the whole set of MECs.  At this point we can distinguish two different approaches towards the reconstruction, i.e. integral and differential.

\subsection{Integral approach}

The formal  solution of  Eq.~(\ref{eq1}) can be written as
\begin{equation} \label{lambda}
\Lambda(t)\equiv\int_0^t dt'\lambda(t')=\ln\left(\frac{\tilde S_j(0)}{\tilde S_j(t)}\right),
\end{equation}
where the suffix $j=1,2$ labels the two components of the vector \begin{equation}
\tilde S(t)=e^{-tM}S(t).\label{S}\end{equation}
We note that in Eq.~(\ref{lambda}) the measurable quantities and the unknown MEC appear on different sides. Analogously, we rewrite Eq.\,(\ref{eq2}) as \begin{equation}
\frac{d}{dt}\tilde X(t)=\tilde D(t),
\label{diff2}\end{equation}
where
\begin{eqnarray}
\tilde X(t) & = & e^{2\Lambda(t)}I_3e^{-t R}X(t), \nonumber \\
\tilde D(t) & = & e^{2\Lambda(t)}I_3e^{-t R}D(t).
\label{XD}\end{eqnarray}
 Eqs. (\ref{S}) and (\ref{XD}) are always invertible, provided one sets the quantity $\sqrt{\delta^2-\omega^2}\equiv\eta$ to the value $i\Omega$ whenever $\eta^2<0$  \cite{Sandulescu1987}.
The formal solution of Eq.~(\ref{diff2}) is given by
$
\tilde X(t)=\tilde X(0)+\int_0^t dt'\tilde D(t'),
$
which, using Eq.~(\ref{XD}),  can be recast in terms of $X(t)$ as
\begin{eqnarray} \label{D}
&&\int_0^t dt'e^{-2\Lambda(t,t')}I_3 e^{(t-t') R}D(t')\nonumber\\&&\qquad\qquad\qquad= X(t)-e^{tR}e^{-2\Lambda(t)}X(0); \quad
\end{eqnarray}
where $\Lambda(t,t')=\int_{t'}^t dt''\lambda(t'')$.
Again, considering $\Lambda(t)$ as a known quantity from   Eq.~(\ref{lambda}), in Eq.~(\ref{D})
 experimentally accessible quantities and MECs are grouped on different sides, respectively right and left-hand. The right-hand sides of both equations can thus be regarded as experimental measurements of the corresponding left-hand sides. The first step is then
to compute the left hand side of Eq.~(\ref{lambda}) using the friction coefficient $\lambda(t)$ provided by the assumed model.  A theoretical value for $\Lambda(t)$, say $\Lambda^{theor}(t)$, is obtained  and its  Fourier transform is performed.
Once obtained the bandwidth associated to $\Lambda^{theor}(t)$ and consequently the amount of points in time required for the full reconstruction (according to the chosen sampling theorem) the following step is to evaluate  $\tilde{S}(t)$ (Eq.~(\ref{S})) on these points by successive applications of the T-C procedure. Finally, the last step is to apply the sampling theorem to the experimental data and to reconstruct the left-hand side of Eq.~(\ref{lambda}), let us denote it by $\Lambda^{expt}(t)$. If the match between $\Lambda^{theor}(t)$ and $\Lambda^{expt}(t)$ is positive within the desired accuracy, one can then proceed further to the reconstruction of the left-hand side of Eq.~(\ref{D}) according to the same procedure.
We note that the integral approach is also feasible in  case of time-dependent Hamiltonian parameters  (i.e. $m(t)$, $\omega(t)$, $\delta(t)$) as long as the generator remains GSP. In general this kind of generators requires a numerical evaluation of  the integrals in Eqs. (\ref{eq1}) and (\ref{eq2}), according to which the suitable sampling theorem must be chosen.

\subsection{Differential approach} \label{meas.der}

The T-C procedure allows to measure not only the cumulants of a given Gaussian state but also their first time derivatives. Indeed,  by measuring each cumulant at two different times $t$ and  $t+\delta t$, its derivative can be estimated via its finite incremental ratio. For instance
\begin{equation}\label{rappIncr}
\frac{d}{dt}\Delta q_t^2\sim\frac{\Delta q_{t+\delta t}^2-\Delta q_t^2}{\delta t},
\end{equation}
where the amount of time $\delta t$ is taken as the smallest time interval which can elapse between two different measurements. Once the cumulants and their derivatives at fixed times are given as experimental inputs,  Eqs.(\ref{eq1}) and (\ref{eq2}) become linear algebraic equations whose unknown are the MECs. Hence  from  Eqs.~(\ref{eq1}) and (\ref{eq2}) we obtain the desired experimentally accessible estimations for the MECs
\begin{eqnarray} \label{est.der}
\lambda^{expt}(t) & \simeq & \delta+\frac{1}{\langle\hat{q}\rangle_t}\left(\frac{1}{m}\langle\hat{p}\rangle_t-\frac{\langle\hat{q}\rangle_{t+\delta t}^2-\langle\hat{q}\rangle_t^2}{\delta t}\right)\nonumber\\
& \simeq & -\delta-\frac{1}{\langle\hat{p}\rangle_t}\left(m\omega^2\langle\hat{q}\rangle_t+\frac{\langle\hat{p}\rangle_{t+\delta t}^2-\langle\hat{p}\rangle_t^2}{\delta t}\right), \qquad \\
D_{qq}^{expt}(t) & \simeq & (\lambda(t)-\delta)\Delta q_t^2-\frac{1}{m}\sigma(q,p)_t \nonumber \\
& & +\frac{\Delta q_{t+\delta t}^2-\Delta q_t^2}{2\delta t}, \\
D_{pp}^{expt}(t) & \simeq & (\lambda(t)+\delta)\Delta p_t^2+m\omega^2\sigma(q,p)_t\nonumber \\
& & +\frac{\Delta p_{t+\delta t}^2-\Delta p_t^2}{2\delta t}, \\
D_{qp}^{expt}(t) & \simeq & \frac{m}{2}\omega^2\Delta q_t^2-\frac{1}{2m}\Delta p_t^2+\lambda(t)\sigma(q,p)_t \nonumber \\
& & +\frac{\sigma(q,p)_{t+\delta t}^2-\sigma(q,p)_t^2}{2\delta t}.\label{est.der2}
\end{eqnarray}
The number of points required to fully reconstruct the MECs depends on the sampling theorem employed.  In Case I, given the prior knowledge of the bandwidth, if the function is band limited or it can be truncated on an effective compact support outside of which the contributions are negligible, the Nyquist-Shannon theorem can be applied. Otherwise one must resort to the additive random sampling theorem, which must be in general employed in Case II.\\
As the integral approach, the differential approach is also suitable in case of time-dependent Hamiltonian parameters as long as the generator remains GSP. The Hamiltonian parameters would then be included among the MEPs to be reconstructed, thus raising their number to seven. Since Eqs.(\ref{eq1}) and (\ref{eq2}) are five, two more equations would be required as for example those for  two higher order cumulants, e.g. ${\rm Tr}(\hat{\rho}(t)\hat q^3)$ and ${\rm Tr}(\hat{\rho}(t)\hat p^3)$. This would not increase the number of experimental measurements, since the higher order moments and cumulants of Gaussian states are completely determined by the first and the second cumulants.\\

{
\subsection{ Comparison between the two approaches} Let us now briefly compare the two procedures described in this section.  The differential approach requires  more
experimental measurements compared to the integral approach, whereas the latter might result more involved from a computational point of
view. For example, within the frame of Case I,
the computation of the first member of
Eq.~(\ref{D}) might  require some numerical or analytical approximations. As any kind of approximation in principle reduces the accuracy of  the reconstruction, the differential approach would be a better choice. On the other hand, if the
computation of the integral functions in Eqs.~(\ref{lambda})-(\ref{D}) does not exhibit
remarkable difficulties, the integral procedure should be preferred, since it
requires a lower number of interactions with the physical system.

In Case II, the integral approach could be employed  to reconstruct the left hand
sides of Eqs.~(\ref{lambda})-(\ref{D}), which are functionals of the unknown MECs. By means
of time derivatives and linear operations on the reconstructed functions,
the MECs can finally be retrieved. However, the time derivatives may amplify
the error of the reconstruction. For instance, the Nyquist-Shannon theorem
 requires a truncation of the Fourier's
frequencies, thus inducing an oscillating behavior of the reconstructed
functions, i. e. introducing the so-called aliasing error (see Appendix A). Even if the oscillations  around the mean (true)
functions are small, the time derivative may increase them. Hence, either one performs a
better reconstruction (e.g. a larger truncation or a random additive
sampling requiring a larger number of measurements) or
 one adopts the differential approach,  thus directly reconstructing the MECs.\\
In general, one could  say that the integral procedure is more advantageous in terms of number of measurements, but requires the ability of solving potentially involved analytical expressions. The differential approach, instead, is more advantageous from the point of view of versatility, as it allows to deal in a straightforward way with complex generators (i.e. exhibiting time-dependent Hamiltonian parameters), at the expenses of a higher number of measurements. Therefore, which of  the two proposed approaches proves better,  strictly depends on the specific case under
investigation.

\section{An example}\label{par:specific case}

To provide an example of  how to implement the proposed integral and differential procedures, in this section we apply both to a specific model i.e. within the perspective of Case I.  As a model to be tested we consider a harmonic oscillator of frequency $\omega$ (our system-particle) linearly coupled to an Ohmic bath of harmonic oscillators. The TIPs of this model are the system-bath coupling constant $\alpha$, the Lorentz-Drude cut-off $\omega_c$ and the temperature $T$ \cite{Petruccione-Breuerlibro2002}.
Starting from a superoperatorial version of the Hu-Paz-Zhang master equation \cite{HPZ1992},
in the weak-coupling limit   (i.e. up to $\alpha^2$) and  using the rotating wave approximation (i.e.  an average over the rapidly oscillating terms),  for this model a generator of the form of Eq.\,(\ref{master equation in q e p}) can be obtained  \cite{Maniscalco2004,Paz2009}.  More specifically, the Hamiltonian parameter $\delta$ is set to 0 and the MECs provided by the model are
 \begin{eqnarray}
\lambda^{theor}(t)&=&\frac{\alpha^2 \omega_c^2 \omega}{ \omega_c^2 + \omega^2} \left\{
    1-e^{-\omega_c t} \left[\cos (\omega t)+ \frac{\omega_c}{\omega} \sin (\omega t)\right]\right\},   \nonumber\\
D^{theor}_{qp}&=&0,\; \frac{m \omega}{\hslash} D^{theor}_{qq}=\frac{D^{theor}_{pp}}{\hslash m \omega}=\frac{\Delta^{theor}(t)}{2},\quad\label{boh}
 \end{eqnarray}
where at high temperature T
  \begin{eqnarray}\label{delta t}
    \Delta^{theor}(t)&=&\frac{ 2 \alpha^2   \omega_c^2  }{\omega_c^2 +\omega^2} \frac{k T}{\hslash}\nonumber \\ & \times & \left\{1-e^{-\omega_c t} \left[\cos (\omega t)- \frac{\omega}{\omega_c}\sin (\omega t)\right]\right\}.\quad
\end{eqnarray}
With this choice of MECs the dissipative generator is  GSP   \cite{Maniscalco2004}.
The Markovian limit is recovered when the time dependent parameters $\lambda^{theor}(t)$ and  $\Delta^{theor}(t)$ reach their stationary values, i.e. at times larger than $1/\omega_c$
\begin{equation}\label{stationary values}
   \lambda^{theor}(t)\rightarrow \frac{\alpha^2 \omega_c^2 \omega}{ \omega_c^2 + \omega^2},\quad \Delta^{theor}(t)\rightarrow \frac{ 2 \alpha^2   \omega_c^2  }{\omega_c^2 +\omega^2} \frac{k T}{\hslash}.
\end{equation}
Usually, when studying quantum Brownian motion, one assumes $\omega_c/\omega \gg 1$ with $\omega_c \rightarrow \infty$, corresponding to a natural Markovian reservoir. In this limit, the thermalization time \cite{Maniscalco2004} is inversely proportional to the coupling strength, while for an out-of-resonance engineered reservoir  with  $\omega_c/\omega \ll 1$ (i.e. highly non Markovian), the thermalization process is slowed down.

Microscopic derivations of Master Equations usually give a time dependent renormalization of Hamiltonian parameters ($m$, $\omega$, $\delta$). However, they are negligible for this benchmark model in the considered limits.

\subsection{Integral approach}

Following the steps of the integral approach we solve Eq.~(\ref{lambda}) using  $\lambda^{theor}(t)$ (Eq.~(\ref{boh})) and we get
\begin{eqnarray} \label{lambda int}
\Lambda^{theor}(t)&=&\frac{\alpha^2  \omega_c^2 \omega^2}{(\omega_c^2+\omega^2)^2}\left\{\omega t \frac{ \omega_c^2 +\omega^2}{\omega^2}-2\frac{\omega_c}{\omega}\right. \nonumber \\
& + & \left.e^{-\omega_c t} \left[2 \frac{\omega_c}{\omega} \cos(\omega t)+\frac{ \omega_c^2 -\omega^2}{\omega^2}
\sin(\omega t)\right]\right\}. \nonumber \\ \end{eqnarray}
The  obtained function does not belong to the functional space $L_2(\mathbb(R))$ thus not matching the condition for the applicability of the  Nyquist-Shannon theorem (see Appendix \ref{sampling}). However, as we are interested in a finite time interval, we can  restrict the support of $\Lambda^{theor}(t)$ to  $[0,\bar t]$  and define:
\begin{equation}
\tilde\Lambda^{theor}(t)\equiv
\begin{cases}
\Lambda^{theor}(t) & t\in[0,\bar t] \\
0 & \mbox{else}
\end{cases}.
\end{equation}
The function $\tilde\Lambda^{theor}(t)$ is in $L_2(\mathbb{R})$ thus can be reconstructed by the Nyquist-Shannon theorem. The discontinuity at $\bar{t}$, by inducing the Gibbs phenomenon (i. e. a finite Fourier sum overshoots at the jump),
 might at this point constitute a source of error. This problem can be anyway overcome by slightly restricting the domain in which the reconstruction of  $\tilde\Lambda^{theor}(t)$ can be trusted to $[0,\bar t-\xi]$ with $\xi>0$.
\begin{figure}[h]
 \centering
\subfigure
{\label{LambdaRM1} \includegraphics[width=0.85\columnwidth]{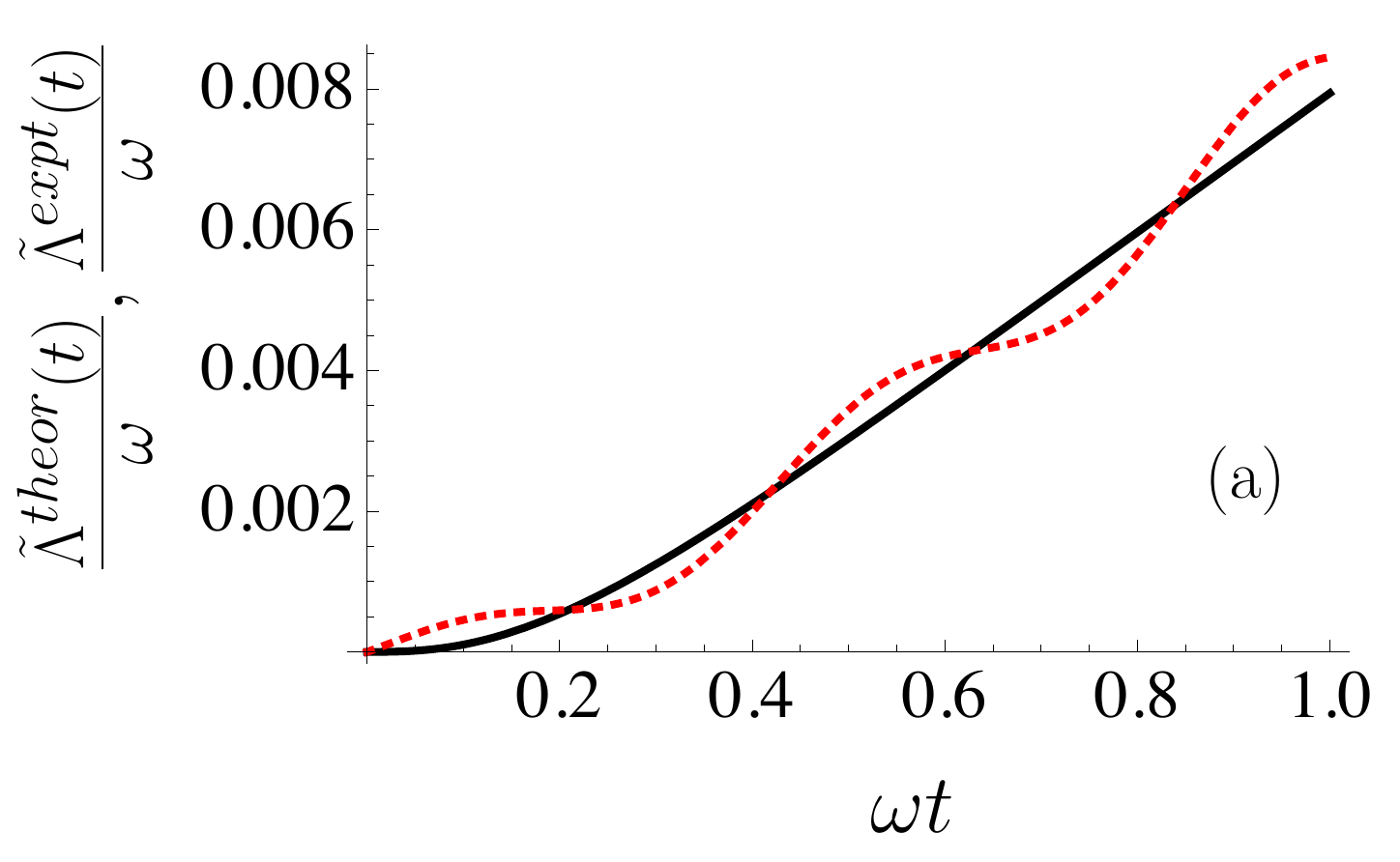}
}
 \hspace{5mm}
 \subfigure
   {\label{LambdaRM2} \includegraphics[width=0.85\columnwidth]{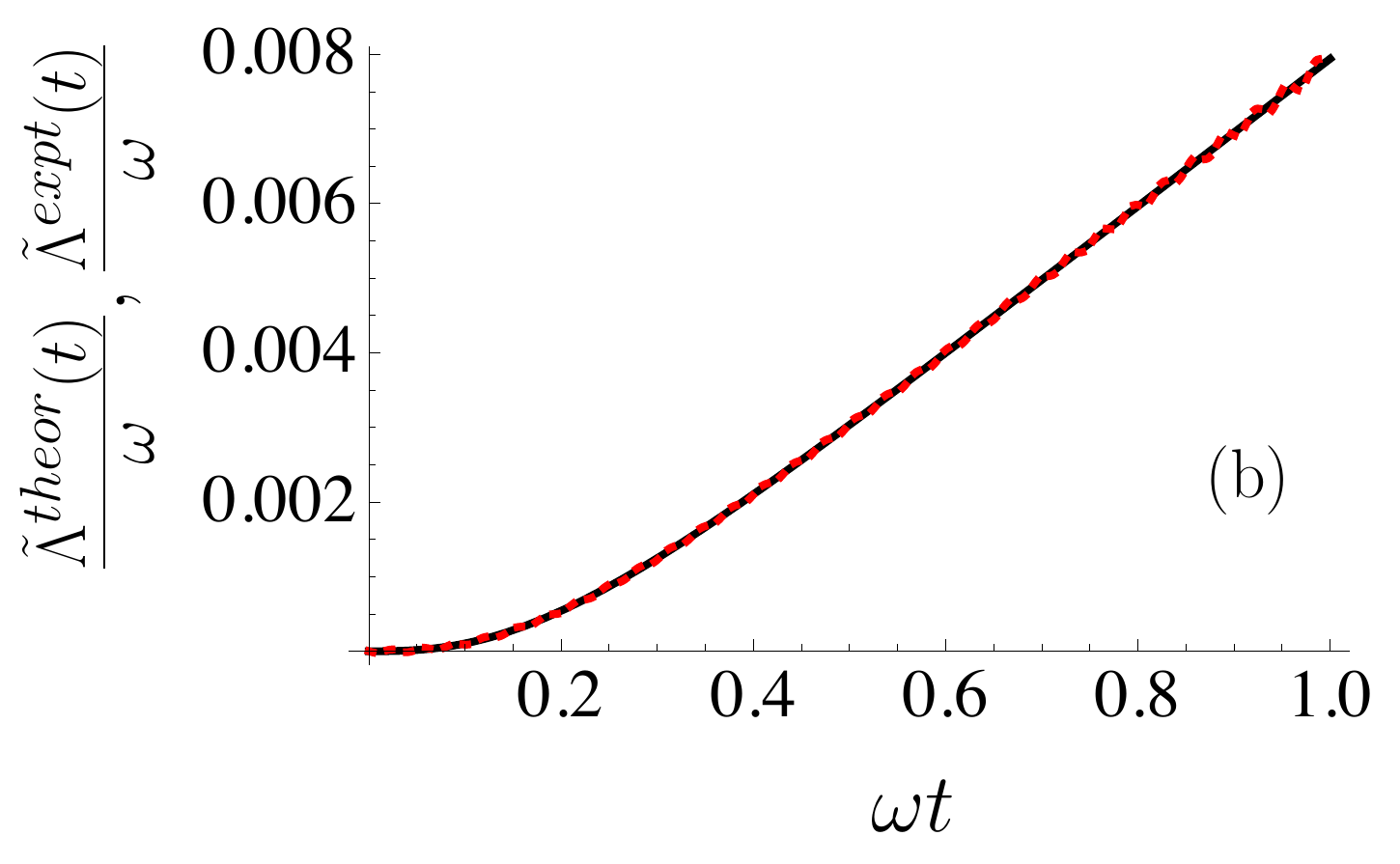}}
\caption{(Color online) Comparison $\tilde{\Lambda}^{expt}(t)$ (dotted)  and i $\tilde{\Lambda}^{theor}(t)$ (continuous), in the Markovian regime, i.e.
$\omega_c/\omega=10$. The support of $\tilde{\Lambda}^{theor}(t)$ is $[0,12/\omega_c]$ and the reconstruction is  trusted in $[0,10/\omega_c]$, with $\alpha = 0.1$ and $T = 10 (\hslash \omega)/k_B$.  In Fig. \ref{LambdaRM1}, contributions smaller than $0.1\%$ of the maximum value of the Fourier transform are neglected (bandwidth  $W=19.4/2\pi$, 7 reconstruction points) while in Fig. \ref{LambdaRM2}, those below $0.01 \%$ (bandwidth $W=196/2\pi$, 74 reconstruction points).}
\label{LambdaRM}
 \end{figure}
\begin{figure}[h]
 \centering
\subfigure
{\label{LambdaNRM1} \includegraphics[width=0.85\columnwidth]{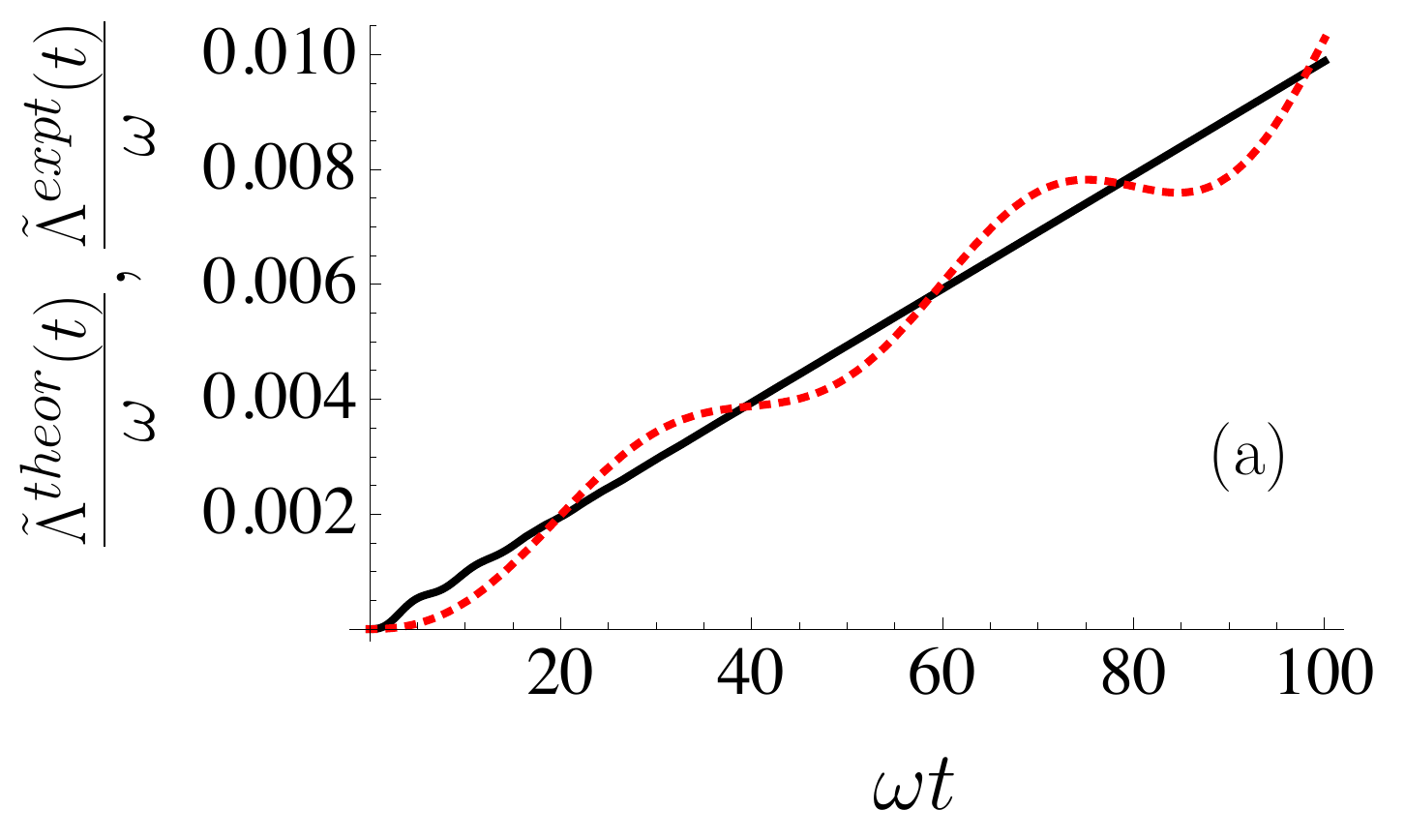}
}
 \hspace{5mm}
 \subfigure
   {\label{LambdaNRM2} \includegraphics[width=0.85\columnwidth]{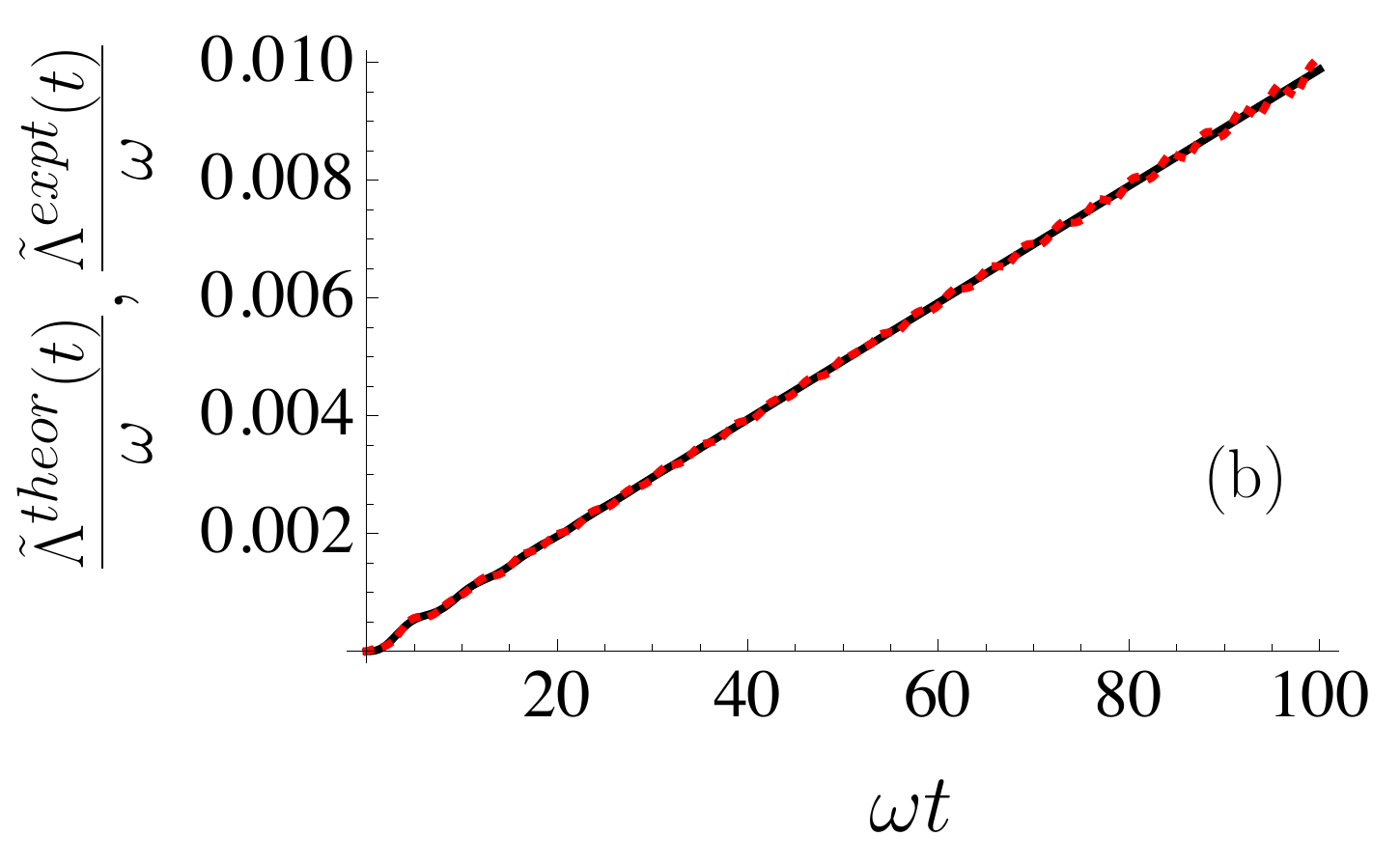}}
   \caption{(Color online) Comparison between $\tilde{\Lambda}^{expt}(t)$ (dotted)  and  $\tilde{\Lambda}^{theor}(t)$ (continuous) in the non-Markovian regime, i.e. $\omega_c/\omega=0.1$.
  The support of $\tilde{\Lambda}^{theor}(t)$ is $[0,12/\omega_c]$, with $\omega_c/\omega=0.1$, and the reconstruction is trusted in $[0,10/\omega_c]$, with  $\alpha = 0.1$ and $T = 10 (\hslash \omega)/k_B$.
  In Fig. \ref{LambdaNRM1}, contributions smaller than $0.1\%$ of the maximum value of the Fourier transform are neglected (bandwidth  $W=0.16/2\pi$, 6 reconstruction points) while in Fig. \ref{LambdaNRM2}, those below $0.01 \%$ (bandwidth $W=1.66/2\pi$, 64 reconstruction points).}
\label{LambdaNRM}
 \end{figure}
Performing the Fourier transform of $\tilde\Lambda^{theor}(t)$ we obtain
\begin{eqnarray}
\label{lambdaf}
&\mathcal{F}&[\tilde\Lambda^{theor}](s)=\frac{\alpha^2\omega\omega_c^2}{(\omega_c^2+\omega^2)^2}
\frac{1}{s^2[\omega^2-\left(s+i\omega_c\right)^2]} \nonumber \\
& \times& \left\{-(\omega_c^2+\omega^2)^2+i e^{\bar t(is-\omega_c)}\left[\left(s+i\omega_c\right)^2-\omega^2\right] \right. \nonumber \\
&\times & \left(i\omega^2+\omega^2s\bar t+i\omega_c^2+\omega_c^2 s\bar t-2\omega_c s\right) \nonumber \\
& +& e^{\bar t(is-\omega_c)}s^2\left(\omega^2+2i\omega_c s-3\omega_c^2\right)\cos(\omega\bar t) \nonumber \\
& +& e^{\bar t(is-\omega_c)}s^2 \left[3\omega\omega_c -\frac{\omega_c^3}{\omega}\left. +i\frac{\omega_c^2-\omega^2}{\omega}s\right]\sin(\omega\bar t)\right\}.\qquad
\end{eqnarray}
From the above equation we note that $\tilde\Lambda^{theor}(t)$ is not band limited, i.e. $\mathcal{F}[\tilde\Lambda^{theor}](s)$ is not defined on a compact support. Nevertheless, since  it is symmetrically decreasing around $s=0$ (modulo a negligible oscillating behavior for higher frequencies), we can define an effective limited bandwidth  and apply the Nyquist-Shannon theorem by truncating its support to a symmetric interval around zero. In Figs. \ref{LambdaRM} and \ref{LambdaNRM} we compare $\tilde{\Lambda}^{theor}(t)$ and $\tilde{\Lambda}^{expt}(t)$ (obtained by applying the reconstruction formula (\ref{reconstruction formula})) in both the Markovian ($\omega_c/\omega=10$) and the non-Markovian ($\omega_c/\omega=0.1$) limit. To respect the conditions of weak coupling and high temperature, in both plots $\alpha$ is set to $0.1$ and $T $ to $10 (\hslash \omega)/k_B$. The reconstruction of  $\tilde{\Lambda}^{expt}(t)$ is trusted in $[0,\bar{t}-\xi=10/\omega_c]$.  As already mentioned, the truncation of the Fourier transform required by the Nyquist-Shannon theorem is the source of the so-called aliasing error in the reconstruction. In order to provide evidence of this phenomenon we compare the results obtained for two different effective supports of the Fourier transform: in Figs. \ref{LambdaRM1} and \ref{LambdaNRM1} we consider only contributions higher than $0.1\%$ of the maximum value, whereas in Figs. \ref{LambdaRM2} and  \ref{LambdaNRM2} those higher than the $0.01\%$.
As expected, the second case returns a better approximation and the aliasing error  becomes negligible, thus not requiring the application of  additive random sampling.
For more involved Fourier transforms, a more general truncation criterion would be taking as effective  support the region where the integral of the Fourier transform is greater than a chosen threshold value.
We skip the reconstruction of the remaining MECs as the procedure is analogous, the only difference being that the integrals on the left-hand side of Eq.~(\ref{D}) require a  numerical evaluation.

\subsection{Differential procedure}

Let us now apply the differential procedure to our model. We note that in this specific example, as the four MECs to be reconstructed effectively reduce to  two ($\lambda(t)$, $\Delta(t)$), we could in principle  allow for time dependent Hamiltonian parameters ($m(t),\omega(t)$) without increasing the number of equations. However, to be consistent with the previous section and with Ref. \cite{Maniscalco2004}, we will consider time-independent Hamiltonian parameters. For our model Eqs.(\ref{est.der}) and (\ref{est.der2}) read
\begin{eqnarray}
\lambda^{expt}(t) & \simeq & \frac{1}{\langle\hat{q}\rangle_t}\left(\frac{1}{m}\langle\hat{p}\rangle_t-\frac{\langle\hat{q}\rangle_{t+\delta t}^2-\langle\hat{q}\rangle_t^2}{\delta t}\right)\nonumber \\
& \simeq & -\frac{1}{\langle\hat{p}\rangle_t}\left(m\omega^2\langle\hat{q}\rangle_t+\frac{\langle\hat{p}\rangle_{t+\delta t}^2-\langle\hat{p}\rangle_t^2}{\delta t}\right), \\
\Delta^{expt}(t) & \simeq & 2\hslash m\omega\lambda(t)\Delta q_t^2-2\hslash\omega\sigma(q,p)_t\nonumber \\
& & +\hslash m\omega\frac{\Delta q_{t+\delta t}^2-\Delta q_t^2}{\delta t} \nonumber \\
& \simeq & \frac{2}{\hslash m\omega}\lambda(t)\Delta p_t^2+\frac{2\omega}{\hslash}\sigma(q,p)_t\nonumber \\
& & +\frac{1}{\hslash m\omega}\frac{\Delta p_{t+\delta t}^2-\Delta p_t^2}{\delta t}.
\end{eqnarray}
 Once again we restrict ourselves to a compact support $[0,\bar{t}]$  (trusting the reconstruction in $[0, \bar{t}-\xi]$) by defining
\begin{eqnarray}\label{restricted functions}
&&\tilde\lambda^{expt}(t)\equiv
\begin{cases}
\lambda^{expt}(t) & t\in[0,\bar t] \\
0 & \mbox{else}
\end{cases},\nonumber\\
&&\tilde\Delta^{expt}(t)\equiv
\begin{cases}
\Delta^{expt}(t) & t\in[0,\bar t] \\
0 & \mbox{else}
\end{cases}.
\end{eqnarray}
The Fourier transforms of the restricted functions read:
\begin{eqnarray}\label{lambapiccoloFourier}
&\mathcal{F}&[ \tilde\lambda^{expt}](s)=\frac{\alpha^2\omega_c^2}{k\left[\omega ^2-(s+i\omega_c)^2\right]\left(\omega ^2+\omega_c^2\right)}\cdot \nonumber \\
& \times& \left\{i\omega\left[\omega^2+\omega_c^2 +e^{(is-\omega_c)\bar t}\left((s+i\omega_c)^2-\omega^2\right)\right]\right. \nonumber \\
& +& e^{\bar t(is-\omega_c)}s\omega(2\omega_c-is)\cos(\bar t\omega) \nonumber \\
& + & \left. e^{\bar t(is-\omega_c)}s\left[\omega_c(\omega_c-is)-\omega^2\right]\sin(\bar t\omega)\right\},
\end{eqnarray}
\begin{eqnarray}\label{DeltaFourier}
&\mathcal{F}&[\tilde\Delta^{expt}](s)=
\frac{2kT\alpha^2\omega_c^2}{s\omega\left[\omega^2-(s+i\omega_c)^2\right]
\left(\omega^2+\omega_c^2\right)}\cdot \nonumber \\
& \times&\left\{i\omega\left[\omega^2+\omega_c^2+e^{\bar t(is-\omega_c)} \left((s+i\omega_c)^2-\omega^2\right)\right]\right. \nonumber \\
&+ & e^{\bar t(is-\omega_c)}s\omega(2\omega_c-is)\cos(\bar t\omega )\nonumber \\
& +& \left.e^{\bar t(is-\omega_c)}s\left[\omega_c(\omega_c-is)-\omega^2\right]\sin(\bar t\omega)\right\}.
\end{eqnarray}
 As in the previous section, the supports of both $\mathcal{F}[\tilde\lambda^{expt}](s)$ and $\mathcal{F}[\tilde\Delta^{expt}](s)$ are not compact, such that we need to truncate them. Having already reconstructed $\tilde\Lambda^{expt}(t)$ in the previous section, we focus here on the remaining MEC. In Figs.\,\ref{DeltaRM} (Markovian regime) and \ref{DeltaRNM} (non-Markovian regime) the reconstructed function, $\tilde{\Delta}^{theor}(t)$, (obtained by applying the reconstruction formula Eq.~(\ref{reconstruction formula})) is compared  with the theoretical curve $\tilde{\Delta}^{theor}(t)$ (Eq.\,(\ref{delta t}))  within the time interval $[0,\bar{t}-\xi=10/\omega_c]$. As in the previous section,  the coupling constant and the temperature are set to $\alpha = 0.1$ and  $T = 10 (\hslash \omega)/k_B$. Again, in order to provide an indication of the sensitivity of the differential procedure to different truncations of the Fourier transform (aliasing error), in Figs. \ref{DeltaRM1} and \ref{DeltaRNM1} the contributions smaller than $0.1 \%$ of its maximum value have been neglected, whereas in Figs. \ref{DeltaRM2} and \ref{DeltaRNM2} those smaller than the
 $0.01 \%$. Analogously to what happened for the integral procedure, the truncation to $0.01 \%$ returns a better reconstruction and the aliasing error becomes almost negligible (thus not requiring additive random sampling).
\begin{figure}
 \centering
\subfigure
{\label{DeltaRM1}\includegraphics[width=0.85\columnwidth]{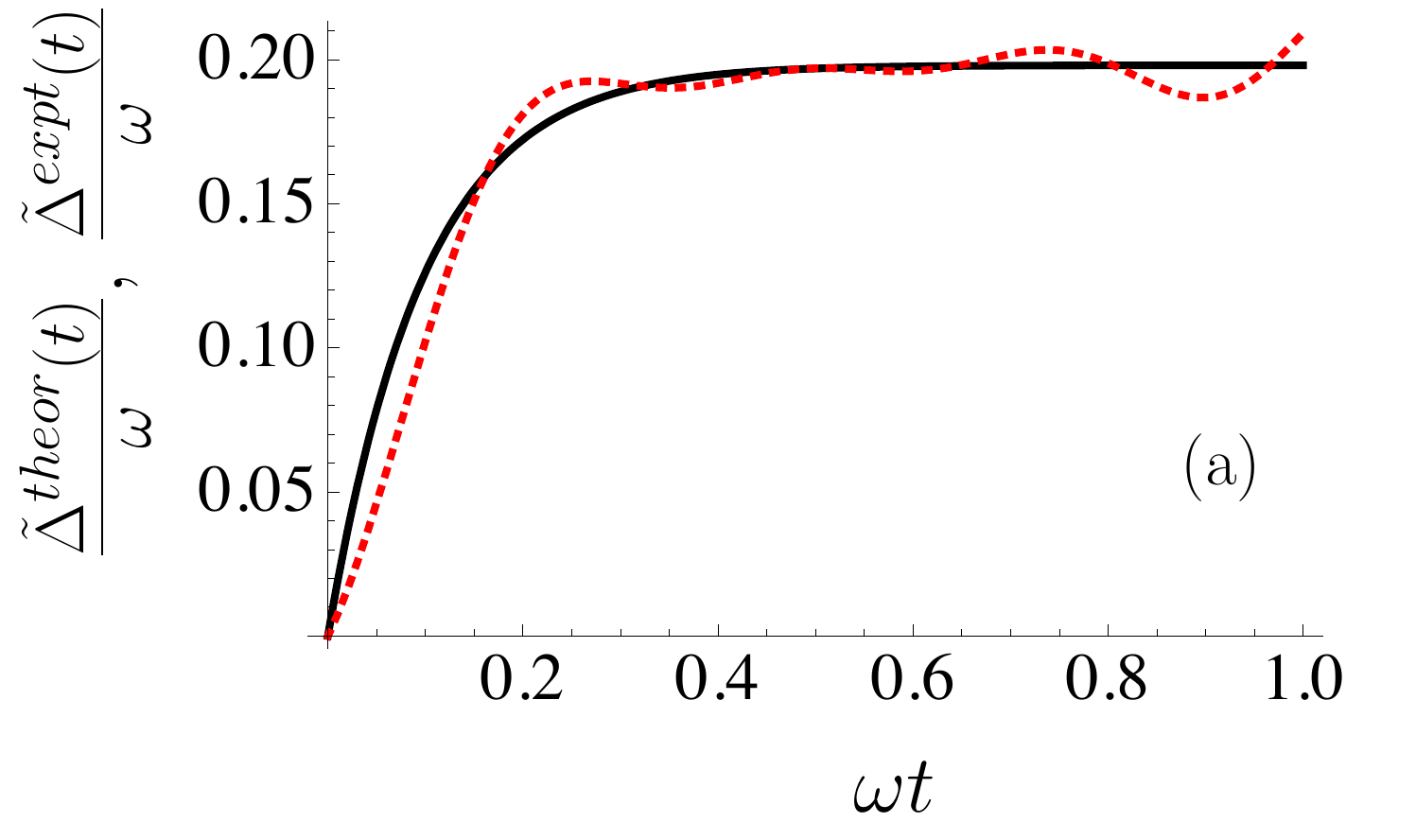}
}
 \hspace{5mm}
 \subfigure
   {\label{DeltaRM2}\includegraphics[width=0.85\columnwidth]{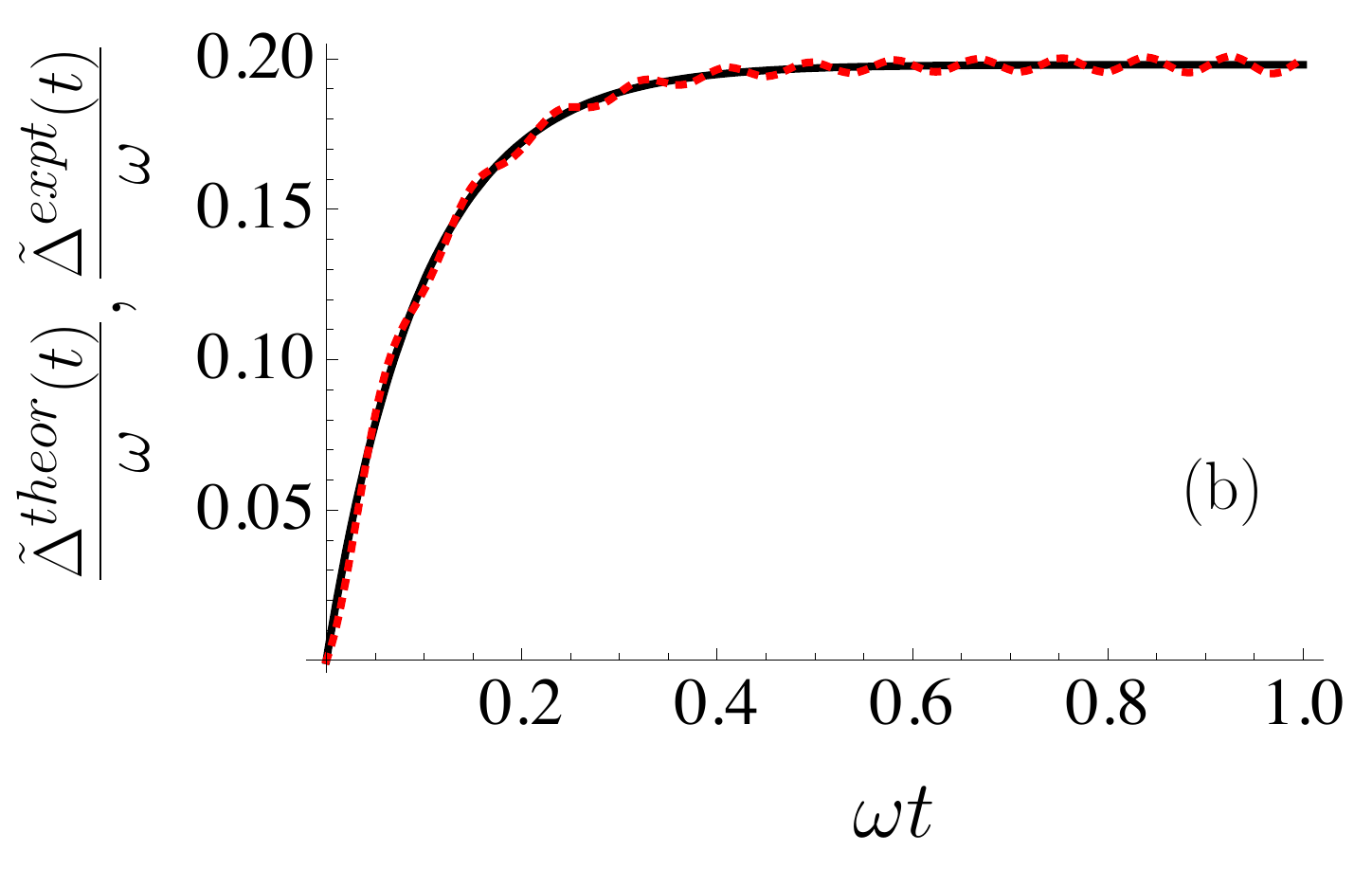}}
\caption{(Color online) Comparison between  $\tilde{\Delta}^{expt}(t)$ (dotted)  and $\tilde{\Delta}^{theor}(t)$ (continuous) in the Markovian regime, i.e. $\omega_c/\omega=10$. The support of $\tilde{\Delta}^{expt}(t)$ is $[0,12/\omega_c]$ and the trusted  interval is $[0,10/\omega_c]$ with  $\alpha = 0.1$ and $T = 10 (\hslash \omega)/k_B$.   In Fig. \ref{DeltaRM1}, contributions smaller than $0.1\%$ of the maximum value of the Fourier transform are neglected (bandwidth  $W=19.5/2\pi$, 7 reconstruction points) while in Fig. \ref{DeltaRM2}, those below $0.01 \%$ (bandwidth $W=73/2\pi$, 34 reconstruction points).}
\label{DeltaRM}
 \end{figure}
\begin{figure}
 \centering
\subfigure
{\label{DeltaRNM1}\includegraphics[width=0.9\columnwidth]{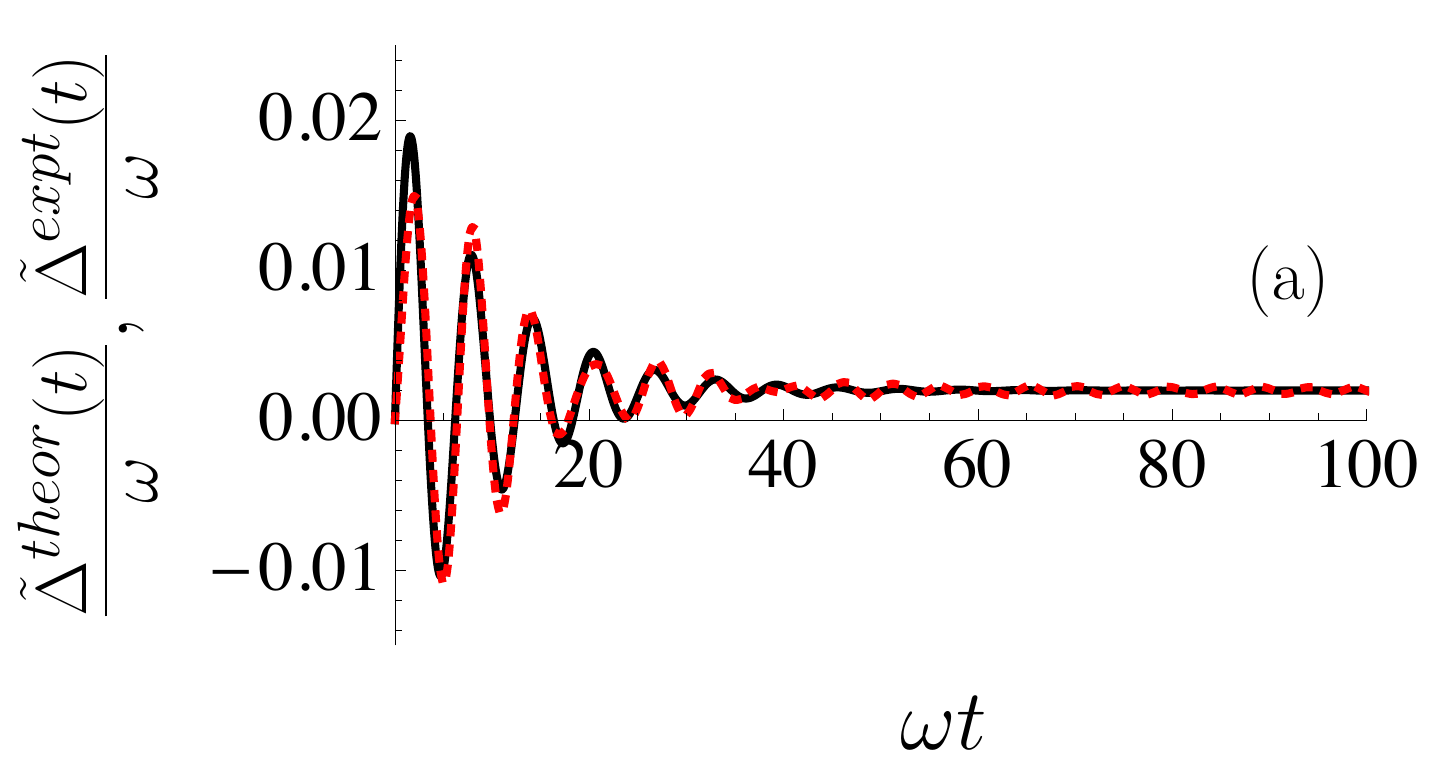}
}
 \hspace{5mm}
 \subfigure
   {\label{DeltaRNM2}\includegraphics[width=0.9\columnwidth]{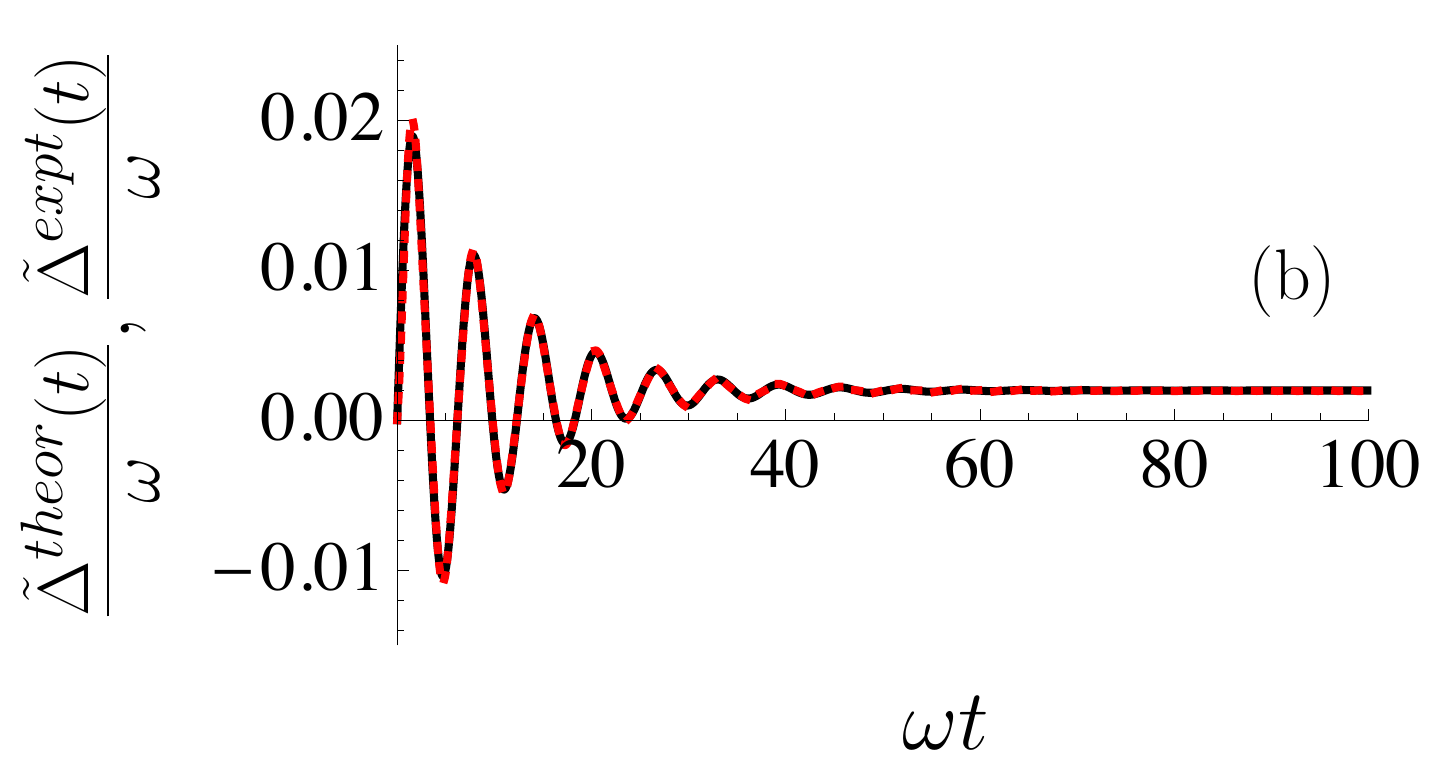}}
\caption { (Color online) Comparison between the reconstructed function $\tilde{\Delta}^{expt}(t)$ (dotted)  and its theoretical behavior $\tilde{\Delta}^{theor}(t)$ (continuous) in non-Markovian regime, i.e. $\omega_c/\omega=0.1$.  The support of $\tilde{\Delta}^{expt}(t)$ is $[0,12/\omega_c]$ and the trusted  interval is $[0,10/\omega_c]$ with  $\alpha = 0.1$ and $T = 10 (\hslash \omega)/k_B$.   In Fig. \ref{DeltaRNM1}, contributions smaller than $0.1\%$ of the maximum value of the Fourier transform are neglected (bandwidth  $W=1.32/2\pi$, 50 reconstruction points) while in Fig. \ref{DeltaRNM2}, those below $0.01 \%$ (bandwidth $W=3/2\pi$, 114 reconstruction points).
\label{DeltaRNM}}
 \end{figure}

\section{Conclusions}\label{par:Summary and Conclusions}

In this paper we have proposed a procedure to reconstruct the master equation coefficients in a non-Markovian scenario by means of a finite and discrete set of tomographic measurements.  In particular, our proposal applies to non-Markovian convolutionless Gaussian Shape Preserving dissipative generators. While the tomogram-to-cumulants procedure provides the local-in-time relation between   the evolved cumulants of a Gaussian probe and experimental measures,  a careful interplay between dynamical equations and sampling theorems tells us how these measurements must be repeated in time to access full global information about the time-dependent master equation coefficients.
We have then proposed two different approaches, integral and differential,  towards full reconstruction,  and provided an explicit example of their application.
According to whether or not one has previous knowledge of the master equation coefficients to be reconstructed, our proposal can be cast within two in principle distinct perspectives. If on one hand
it provides a tool to gain information about fully unknown quantities, on the other, if any {\it a priori} knowledge is available, it represents a consistency test of the theoretical assumptions   associated to a given model, therefore providing reliable information about the overlap between the real and the assumed dissipative dynamics. In case one has previous knowledge of the MECs and can perform the intergrals in Eqs. (\ref{lambda}),(\ref{D}), the integral approach proves more efficient in terms of the amount of required measurements, otherwise the differential approach represents the most appropriate one even though it requires more measurements.
 Furthermore, along this line of thought our procedure might also prove useful  to test the theoretical predictions associated to a given model, such as the crossing of the Lindblad-non Lindblad border investigated in \cite{Maniscalco2004}.
 Our proposal opens up several interesting questions which are going to be the subject of further future investigation.  In facts  how our approach can be recast within an estimation theory perspective, along the lines of \cite{Alex2007}, represents a relevant open scenario. Another relevant point  to investigate is whether the proposed protocol can be enhanced by employing entangled Gaussian states as a probe. Finally,
 whether or not our proposal can be further generalized and employed in presence of memory kernels represents a non trivial question. Indeed, finding a way to reconstruct with a finite amount of measurements the coefficients of non-Markovian dissipative generators with memory represents both a highly challenging and interesting task which deserves further future investigation.

\section{Acknowledgements}
 We warmly thank P. Facchi for many interesting and useful discussions. We thank as well E. Milotti for useful discussions on the sampling theorem. A. D. P. acknowledges the financial support of the European Union through the Integrated Project EuroSQIP.
\appendix
\section{Sampling theorems} \label{sampling}

Here we provide some details about the Nyquist-Shannon and the additive random sampling theorems,
for a full review on this topic  see \cite{Jerry1977,Unser2000}. The simpler sampling theorem, known as Nyquist-Shannon theorem, deals with functions whose Fourier transform has a compact support. The theorem allows these functions to be reconstructed starting from a discrete and infinite set of values. The theorem is stated as follows:\\
\noindent
\textbf{Theorem 1.} (Nyquist-Shannon theorem)\quad
\textit{If a function $F(t)$ has no frequency higher than $W$, that is the support of its Fourier transform is contained in $[-2\pi W,2\pi W]$, then it is completely determined by giving its ordinates at a series of points spaced $\frac{1}{2W}$ apart.}

\medskip

The reconstruction formula is
\begin{equation}\label{reconstruction formula}
F(t)=\sum_{n=-\infty}^\infty F\left(\frac{n}{2W}\right)\frac{\sin\pi(2Wt-n)}{\pi(2Wt-n)}.
\end{equation}
The most frequent sources of error  are the truncation error, the aliasing error, the round-off error and the jittering error. The truncation error arises from considering a finite sampling instead of an infinite one, as in practice it is unfeasible to sample and store an infinite number  of values (unless some regular behavior of the function to reconstruct can be postulated). The aliasing error occurs whenever a function which is not band-limited is reconstructed by means of procedures suitable for band-limited functions which is often the case  since band-limited functions are very peculiar. Whenever a truncation or an aliasing error occurs, it means that there are different functions matching the exploited sampling, anyway there exist conditions to bind this kind of errors \cite{Jerry1977,Unser2000}. Whereas the two previous errors are sampling-related, whereas the round-off and the jittering error are linked to the precision of the experimental apparatus. In particular, the round-off error is caused by errors affecting the sampling values and, finally, the jittering error is due to errors affecting the sampling times.  \\

A remarkable generalization of the sampling theorem is developed in \cite{Shapiro1960,Beutler1970} and requires an additive random sampling. The function is sampled at additively randomly chosen points $t_n=t_{n-1}+\gamma_n$, where $\{\gamma_n\}$ is a family of independent identically distributed random variables, whose probability distribution $p(t)$ obeys the following constraints
\begin{equation}
p(t)\in\mbox{L}^2(\mathbb{R}),\quad p(t)=0\quad\mbox{for}\quad t<0,\quad \mathbb{E}[\gamma_n]=h<\infty,\label{p}
\end{equation}
where $\mathbb{E}[\cdot]$ is  the expected value and $h$ is the average spacing of the sampling. Additive random samplings may be alias-free, namely they may provide reconstructions without any aliasing error. As a consequence, non band-limited functions may as well be reconstructed. Examples of alias-free samplings are discussed in \cite{Shapiro1960,Beutler1970}. The following theorem provides the condition for an additive sampling to be alias-free:\\
\noindent
\textbf{Theorem 2.}\quad
\textit{An additive random sampling is alias-free if the characteristic function $\phi(\omega)=\mathbb{E}[e^{i\omega t}]$ takes no values more than once on the real axis. \\
\noindent
Conversely, if the characteristic function $\phi(\omega)$ takes the same value at two different points of the open upper half-plane, then aliasing occurs with an additive random sampling.}
\medskip

The reconstruction formula is a bit involved and is given in \cite{Shapiro1960,Beutler1970}. We emphasize that in this case the input sampling is  not a countable set of values but  rather  a countable set of averages as the sample spacing is a random variable itself and we make an average over all possible spacings. If the spacing distribution is highly picked on $h$, i.e. $p(t)=\delta(t-h)$, the spacing variance  vanishes $\mathbb{E}[\gamma_n^2]=\mathbb{E}^2[\gamma_n]$, and the reconstruction procedure recovers the usual one with  equally spaced samplings. If the spacing variance is comparable with its mean value $\mathbb{E}[\gamma_n^2]-\mathbb{E}^2[\gamma_n]\sim h^2$, then all spacings are highly probable and we get an effective continuous sampling. An unavoidable source of error  in the implementation of this sampling theorem is that it involves  functions defined on arbitrarily small times  \cite{Shapiro1960,Beutler1970} whereas any experimental apparatus exhibits a dead working time interval to record and process data.
Finally we note that we want to reconstruct functions  defined only for $t>0$ hence we do not encounter any lower truncation error. The upper truncation error can be also avoided by reconstructing slightly different functions: $F(t)(1-\theta(t-\bar t))$ instead of $F(t)$. The differences arise from the integral transforms involved in the reconstruction formula, but if we are interested in reconstructing functions in the experimentally accessible timescales, values greater than the threshold $\bar t$ of experimentally reachable times can be neglected. The same trick can be exploited to avoid errors due to a finite input size  for an additive random sampling. Indeed if the sampling probability is well localized, each $p_n(t)$ is  localized as well. The probabilities $p_n(t)$ localized at  times larger than the threshold $\bar t$ do not contribute and the corresponding $f_n$ vanish.


\end{document}